# Reducing laser beam fluence and intensity fluctuations in symmetric and asymmetric compressors

**Efim Khazanov**

*Gaponov-Grekhov Institute of Applied Physics of the Russian Academy of Sciences*

**Abstract** All space-time coupling effects arising in an asymmetric optical compressor [1] consisting of two non-identical pairs of diffraction gratings are described analytically. In each pair, the gratings are identical and parallel to each other, whereas the distance between the gratings, the groove density, and the angle of incidence are different in different pairs. It is shown that the compressor asymmetry does not affect the far field fluence and on-axis focal intensity. The main distinctive feature of the asymmetric compressor is spatial noise lagging behind or overtaking the main pulse in proportion to the transverse wave vector. This results in a degraded contrast but reduces beam fluence fluctuations at the compressor output. Exact expressions are obtained for the spectrum of fluence fluctuations and fluence rms that depends only on one parameter characterizing compressor asymmetry. The efficiency of small-scale self-focusing suppression at subsequent pulse post-compression is estimated.

## I. INTRODUCTION

The present-day high-power femtosecond lasers [2] work at a fluence tens of percent lower than laser damage threshold, as spatial noise inevitably leads to both intensity and fluence fluctuations.The spectrum of spatial noise covers scales from several wavelengths to the beam size, which is  tens of centimeters in ultra-high-power lasers. The noise increasing during the propagation in a medium with cubic nonlinearity due to small-scale self-focusing, also known as the Bespalov-Talanov filamentary instability [3], plays an important role in this spectrum. The most hazardous spatial scale that leads to the so-called hot spots is an order of 30 microns [4, 5]. This noise restricts the use of post-compression after a diffraction grating compressor [5, 6], as well as any other transmissive optical elements (frequency doublers, quarter-wave plates, polarizers, beam splitters). All these elements greatly expand the applicability range of ultra-high-power lasers. To suppress filamentation instability, it is necessary to decrease the noise field *intensity* at the time of the maximum of the main pulse (maximum of nonlinearity).

Even if transmissive optics is not used after the compressor, the maximum achievable energy of the output pulse is limited by the breakdown threshold of diffraction gratings, which is especially true for projects of 100 PW lasers proposed in China [7, 8], the USA [9, 10], Japan [11-13], and Russia [14] (see also the review papers [2, 15]). The weakest link is the last, fourth grating, since the breakdown threshold of a femtosecond pulse is much lower than the breakdown threshold of a nanosecond pulse [16], e.g., 228 mJ/cm$^2$ versus 600 mJ/cm$^2$ [17]. Reliable and safe operation of the compressor demands the *fluence* at the fourth grating to be less than the threshold with some margin. The required margin depends on fluence fluctuations. The smaller the fluctuations, the less the margin and the higher the maximum energy and laser power. Fluence fluctuations are determined by the entire spectrum of spatial noise. Note that the standard deviation of the fluence is determined by the integral of the fluctuation spectrum to which low-frequency noise makes a more significant contribution. At the same time, the magnitude of the maximum overshoot depends significantly on high-frequency noise [18].

Thus, it is highly important to reduce both, the fluctuations of laser beam *intensity* and *fluence*. To do this, high-power femtosecond lasers use spatial [4, 19, 20] and temporal [6, 21] self-filtering at free propagation in vacuum. The term "self-filtering" is used to emphasize that no devices are required for this, only propagation in a vacuum over a certain length. Physically, self-filtering is explained by the fact that spatial noise propagates at an angle to the wave vector of the principal wave. That is why the noise lags behind the pulse of the principal wave and walks off the aperture of the principal wave beam, thus leading to temporal and spatial self-filtering, respectively. The detailed theoretical and experimental studies carried out in [22, 23] showed that free space acts as a spectral filter of fluence fluctuations, with the filter transmittance being equal to the intensity autocorrelation function for spatial self-filtering and to the square of the field autocorrelation function for the temporal one.

It is obvious that the compressor, in which the beam passes a considerable distance, is also a fluence fluctuation filter, but the properties of this filter have not been studied before. Moreover, to smooth fluence fluctuations it was recently proposed to use a pair of prisms [24], an asymmetric four-grating compressor (AFGC) [1], as well as a compressor with one pair of gratings [25, 26], which is a particular case of the AFGC. Numerical simulations were carried out in [1, 25, 26] but no analytical theory was constructed.

The purpose of this paper is to provide an analytical theory aimed at quantifying the smoothing efficiency of intensity and fluence in both symmetric and asymmetric compressors and to compare self-filtering and AFGC filtering.

## II. THE FIELD AT THE COMPRESSOR OUTPUT

We will consider a compressor consisting of two pairs of parallel gratings (Fig. 1), on which a laser beam with an arbitrary temporal and spatial spectrum is incident. The z axis coincides with the beam trajectory at the compressor input with $k_{x,y} = 0$ and $\omega = \omega_0 = ck_0$, where $\omega_0$ and $k_0$ are, respectively, the carrier frequency and wave vector. The angles $\alpha_{1,2}$ and $\beta_{1,2}$ of this beam are related by the grating expression as

$$sin\beta_{1,2} = m \frac{2\pi}{k_z(\omega_0)} N_{1,2} + \sin \alpha_{1,2}, \qquad (1)$$

where $N_{1,2}$ is the groove density, $m$ is the diffraction order, and $k_z^2 = k_0^2 - k_x^2 - k_y^2$. Hereinafter, the subscripts "1", "2" refer to the first and second pair of gratings, respectively. The magnitudes of $\alpha, N$ and $L$ in the second pair of gratings may be either like in the first pair or different from them. In the first case, the compressor will be called a symmetric or Treacy compressor (TC) [27]. In the second case, it will be called an asymmetric compressor or AFGC [1].

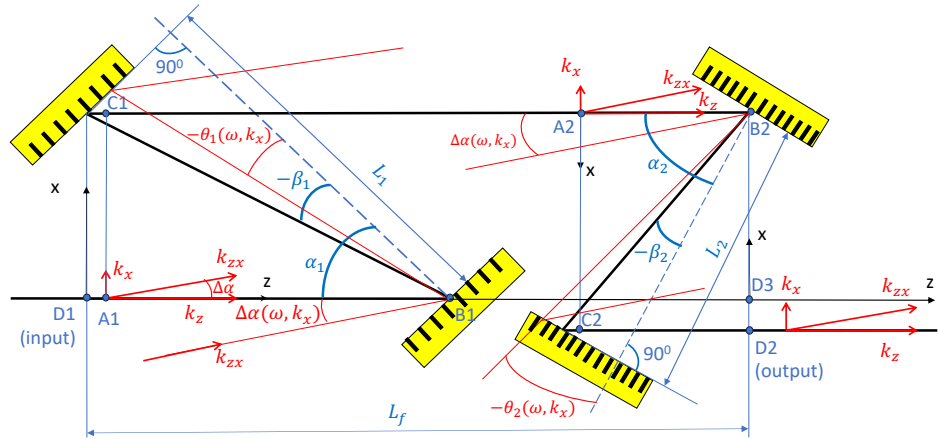

Fig. 1. AFGC scheme. Gratings in each pair are parallel and identical. Distances between the gratings $L$, groove densities $N$ and incidence angles $\alpha$ are different in different pairs.

The ray with arbitrary $k_x$ and $\omega$ in Fig. 1 is shown by the red color. The angle of incidence on the first grating for this ray is $\alpha + \Delta\alpha(\omega, k_x)$, where $\Delta\alpha = \text{atan}\left(\frac{k_x}{k_z(\omega)}\right)$. The angle of the ray reflection from the grating is $\theta(\omega, k_x)$. It depends on $k_x$ and $\omega$, with $\theta(\omega = \omega_0, k_x = 0) = \beta$. The angles of reflection $\theta$ and $\beta$ are counted to the right of the normal (i.e. for a mirror, $\beta = \alpha$). In the minus first diffraction order ($m = -1$), the angles $\theta$ and $\beta$ are negative: $\beta < 0$, $\theta < 0$; therefore, the minus sign is indicated in the figure.

Let the real electric field at the compressor input at the point D1 be $\mathcal{E}_{in}(t, x, y) = Re\{E_{in}(t, x, y)e^{i\omega_0 t - ik_0 z}\}$, where $E_{in}(t, x, y)$ is the complex amplitude. The relationship between $E_{in}(t, x, y)$ and the complex amplitude at the point D2 at the compressor output $E_{out}(t, x, y)$ is more readily found in the spectral representation. Hereinafter, the spectra will be designated by the same characters but with different arguments:

$$E_{in,out}(\omega, k_x, k_y) = \mathbf{F}\{E_{in,out}(t, x, y)\} \quad E_{in,out}(t, x, y) = \mathbf{F}^{-1}\{E_{in,out}(\omega, k_x, k_y)\}, \tag{2}$$

where $\mathbf{F}$ and $\mathbf{F}^{-1}$ are the forward and inverse 3D Fourier transforms. Our consideration is restricted to the case of parallel and identical gratings in each pair. This ensures that the input and output wave vectors are parallel. In other words, the input and output wave vectors are always equal, so are the frequencies. Hence, $E_{out}$ and $E_{in}$ are related by

$$E_{out}(\omega, k_x, k_y) = \exp\left(i\Psi(\omega, k_x, k_y)\right) \cdot E_{in}(\omega, k_x, k_y) \tag{3}$$

$$E_{out}(t, x, y) = \mathbf{F}^{-1}\left\{\exp\left(i\Psi(\omega, k_x, k_y)\right) \cdot \mathbf{F}\{E_{in}(t, x, y)\}\right\}, \tag{4}$$

where $\Psi(\omega, k_x, k_y)$ is the phase incursion between the input and output planes. For simplicity, in (3) it is assumed that there are no losses and the absolute value of the transmission coefficient is equal to unity. Before passing to the expression for $\Psi(\omega, k_x, k_y)$, in important general note should be made that follows directly from (3) and Parseval's theorem. The compressor changes the near field distribution $E_{out}(\omega, k_x, k_y)$, $E_{out}(t, k_x, k_y)$, but *in no way affects the fluence* in the far field:

$$\int |E_{out}(t, k_x, k_y)|^2 dt = \int |E_{in}(t, k_x, k_y)|^2 dt$$

This result was obtained numerically for specific examples [1, 25]. We showed analytically that this statement is true for any compressor, independent of its parameters. Analogously it can be proved that the intensity on the beam axis in the far field $|E_{out}(t, k_x = k_y = 0)|^2$ does not depend on the compressor symmetry either. The only necessary condition for this is parallel and identical gratings in each pair, otherwise (3) does not hold.

The expression for $\Psi(\omega, k_x, k_y)$ (the derivation is presented in Appendix) has the form

$$\Psi(\omega, k_x, k_y) = L_1 \frac{\omega}{c}\left(\cos\theta_1 + \cos\left\{\alpha_1 + \frac{k_x c}{\omega}\right\}\right) + L_2 \frac{\omega}{c}\left(\cos\theta_2 + \cos\left\{\alpha_2 - \frac{k_x c}{\omega}\right\}\right) + L_f \frac{\omega}{c} - \frac{c}{2\omega}k_x^2 L_f - \frac{c}{2\omega}k_y^2[L_f + (\cos\{\theta_1(k_{x,y} = 0)\} + \cos\alpha_1)L_1 + (\cos\{\theta_2(k_{x,y} = 0)\} + \cos\alpha_2)L_2], \tag{5}$$

where

$$\sin\theta_{1,2}(\omega, k_x, k_y) = m\frac{2\pi c}{\omega}N_{1,2}\left(1 + \frac{c^2}{2\omega^2}k_y^2\right) + \sin\left\{\alpha_{1,2} \pm \frac{k_x c}{\omega}\right\} \tag{6}$$

Here, $L_{1,2}$ is the distance between the gratings along the normal, and $L_f$ is the distance between the input and output planes of the compressor (Fig. 1). The first two terms in (5) depend, among others, on the first power of $k_x$. For TC, i.e. for $L_2 = L_1$; $\alpha_2 = \alpha_1$; $N_2 = N_1$; $\theta_2(k_x) = \theta_1(-k_x)$, this dependence disappears, while it remains for AFGC. This is the difference between AFGC and TC. A compressor with a single-grating pair (single-pass single-grating pair SSGC) [25] is a particular case of an AFGC with $L_2 = 0$.

The expression (4) with allowance for (5, 6) is sufficient for numerical simulation. To continue the analytical analysis, we will expand (5) in a Taylor series.

### III. EXPANDING $\Psi(\omega, k_x, k_y)$ IN A TAYLOR SERIES

By expanding (5) in a Taylor series with respect to $k_x, k_y$ and $\Omega = \omega - \omega_0$ we obtain

$$\Psi(\omega, k_x, k_y) = \psi''_{x\omega}\Omega k_x + \frac{1}{2}\psi'''_{x\omega\omega}\Omega^2 k_x + \frac{1}{2}\psi''_{xx}k_x^2 + \frac{1}{2}\psi''_{yy}k_y^2 + \frac{1}{2}\psi'''_{xx\omega}\Omega k_x^2 + \frac{1}{2}\psi'''_{yy\omega}\Omega k_y^2, \quad (7)$$

where

$$\psi'_{x\omega} = -\frac{1}{\omega_0}(A_1 L_1 - A_2 L_2) \qquad A = \frac{\cos\alpha}{\cos^3\beta}(\sin\alpha - \sin\beta) \qquad (8)$$

$$\frac{1}{2}\cdot\psi'''_{x\omega\omega} = \frac{1}{\omega_0^2}(B_1 L_1 - B_2 L_2) \qquad B = A\left(1 - \frac{3}{2}\sin\beta\frac{\sin\alpha - \sin\beta}{\cos^2\beta}\right) \qquad (9)$$

$$\frac{1}{2}\cdot\psi''_{xx} = -\frac{c}{\omega_0}\frac{1}{2}(C_1 L_1 + C_2 L_2 + L_f) \qquad C = \frac{\cos^2\alpha}{\cos^3\beta} - tg\beta\sin\alpha + \cos\alpha \qquad (10)$$

$$\frac{1}{2}\cdot\psi''_{yy} = -\frac{c}{\omega_0}\frac{1}{2}(D_1 L_1 + D_2 L_2 + L_f) \qquad D = \frac{1 + \cos(\alpha + \beta)}{\cos\beta} \qquad (11)$$

$$\frac{1}{2}\cdot\psi'''_{xx\omega} = \frac{c}{\omega_0^2}\frac{1}{2}(E_1 L_1 + E_2 L_2 + L_f) \qquad E = C + \frac{\sin\alpha - \sin\beta}{\cos^3\beta}\left(\sin\alpha - 3\sin\beta\frac{\cos^2\alpha}{\cos^2\beta}\right) \qquad (12)$$

$$\frac{1}{2}\cdot\psi'''_{yy\omega} = \frac{c}{\omega_0^2}\frac{1}{2}(F_1 L_1 + F_2 L_2 + L_f) \qquad F = \frac{\cos 2\beta + \cos(\beta - \alpha)}{\cos\beta} - \frac{(\sin\beta - \sin\alpha)^2}{\cos^3\beta} \qquad (13)$$

$$A_{1,2} = A(\alpha_{1,2}, \beta_{1,2}); \; B, C, D, E, F \text{ -- analogously} \qquad (14)$$

The lower-case letter $\psi$ designates the values of the corresponding derivatives for $k_x = k_y = 0; \; \omega = \omega_0$, for example:

$$\psi''_{x\omega} = \frac{\partial^2 \Psi(\omega, k_x, k_y)}{\partial k_x \partial \omega}\bigg|_{k_{x,y} = 0; \; \omega = \omega_0} \qquad (15)$$

Some terms are omitted in (7). These include terms with the first derivative with respect to $k_y$ that are equal to zero, since $\Psi$ depends only on $k_y^2$. Next, $\Psi(\omega = \omega_0, k_x = k_y = 0)$ is a constant that may also be put equal to zero. We omit $\psi'_\omega \Omega$ that is the time delay which does not depend on $k_{x,y}$ and $\psi'_x k_x$ that is the $\Omega$-independent shift along the x axis. These terms are equivalent to the shift of the origin of the $t$ and $x$ axes which is of no importance. Also, we omit all the terms containing derivatives only with respect to $\omega$: $\frac{1}{2}\psi''_{\omega\omega}\Omega^2$, $\frac{1}{6}\psi'''_{\omega\omega\omega}\Omega^3$, and so on. These terms correspond to different degrees of time dispersion for a field with $k_x = 0$. In this work aimed at studying the space-time effects, we will omit the above mentioned terms and will assume that the dispersion introduced by the compressor corresponds to the pulse dispersion at its input. Therefore, for the zero spatial frequency $k_x = 0$, the output pulse is Fourier-transform-limited. Finally, we neglect all terms with derivatives with respect to $k_{x,y}$ higher than the second one, as well as all derivatives higher than the third one in view of their smallness.

Let us now address the remaining six terms in (7) and the respective physical effects. The first two terms are proportional to the first power of $k_x$: $\psi''_{x\omega}\Omega k_x$ is the time lag by $\psi''_{x\omega}k_x$ or the shift along the x axis by distance $\psi''_{x\omega}\Omega$; $\frac{1}{2}\psi'''_{x\omega\omega}\Omega^2 k_x$ is the pulse stretching ($GVD = \frac{1}{2}\psi'''_{x\omega\omega}k_x$) or the shift along the x axis by distance $\frac{1}{2}\psi'''_{x\omega\omega}\Omega^2$. As seen from (8, 9), these terms in the TC are equal to zero. The second two terms in (7), $\frac{1}{2}\psi''_{xx}k_x^2$ and $\frac{1}{2}\psi''_{yy}k_y^2$, correspond to diffraction that is different along the x and y axes, as well as to spatial self-filtering, i.e. to the shift along the x and y axes by distance $\frac{1}{2}\psi''_{yy}k_{x,y}$. Finally, the last two terms in (7) correspond to temporal self-filtering that is the time lag by $\frac{1}{2}\psi'''_{xx\omega,yy\omega}k_{x,y}^2$. Note that the last four terms and the corresponding effects occur not only in a compressor but also in free space. Unlike the case of the compressor, in free space they are isotropic. The expressions in parentheses for $\psi$ in (10, 11, 12, 13) may be interpreted as the effective compressor length in terms of the corresponding effect. For example, $(C_1 L_1 + C_2 L_2 + L_f)$ is the effective length in terms of diffraction along the x axis, and $(F_1 L_1 + F_2 L_2 + L_f)$ is the effective length in terms of temporal self-filtering along the y axis. Values of the $A-F$ constants for compressors borrowed from some works are given by way of example in Table 1. Let's go into the details of these effects in different compressors.

Table 1. Parameters of the compressors: TC [14], AFGC [1] and SSGC [25].

|  | TC | AFGC | | SSGC |
|---|---|---|---|---|
| $\alpha$, degrees | 45.5 | 61.0 | 61.0 | 57.0 |
| $N$, 1/mm | 1200 | 1400 | 1400 | 1400 |
| $L$, mm | 1850 | 1401 | 1079 | 2300 |
| $A$ | 0.97 | 0.84 | 0.84 | 1.00 |
| $B$ | 1.66 | 1.67 | 1.67 | 2.12 |
| $C$ | 1.61 | 1.20 | 1.20 | 1.40 |
| $D$ | 2.07 | 1.99 | 1.99 | 2.10 |
| $E$ | 3.49 | 3.35 | 3.35 | 3.88 |
| $F$ | -0.32 | -1.45 | -1.45 | -1.61 |
| $\Delta L$, mm | 0 | 271 | | 2303 |
| $\Delta L'$, mm | 0 | 539 | | 4881 |

## IV. TC AS A FILTER OF SPATIAL FREQUENCIES

The derivatives (8, 9) vanish in the TC, and only two effects remain in (7) (spatial self-filtering and temporal self-filtering that are proportional to $k_{x,y}^2$. Thus, from the point of view of spatial effects, the TC is similar to free space, but it introduces astigmatism, because the x and y axes are no longer equivalent. It is seen from (10, 11) that, in terms of spatial self-filtering and diffraction, TC supplements the free space length $L_f$ with an additional length that is different for the x and y axes: $2C_1 L_1$ for the x axis and $2D_1 L_1$ for the y axis. With this astigmatism taken into account, the compressor is equivalent to free space with different lengths on the x and y axes:

$$L_{sx} = 2C_1 L_1 + L_f \quad L_{sy} = 2D_1 L_1 + L_f \tag{16}$$

Thus, the TC is an anisotropic transmissive filter. The transmission coefficient of such a filter can be obtained by generalizing expressions from [22, 23] to the anisotropic case.

In free space, the time lag $\tau$ of the pulse propagating at an angle to the z axis is proportional to $(k_x^2 + k_y^2)$. In a symmetric compressor, the coefficients of proportionality for $k_x^2$ and $k_y^2$ are not the same, see Fig. 2. From (12, 13) we obtain

$$\tau(k_x, k_y) = \frac{1}{2}\psi'''_{xx\omega}k_x^2 + \frac{1}{2}\psi'''_{yy\omega}k_y^2 = \frac{L_{tx}}{2c}\frac{k_x^2}{k_0^2} + \frac{L_{ty}}{2c}\frac{k_y^2}{k_0^2}, \tag{17}$$

where

$$L_{tx} = 2E_1L_1 + L_f \qquad\qquad L_{ty} = 2F_1L_1 + L_f \qquad (18)$$

Thus, from the point of view of temporal self-filtering, the TC adds to the free space length $L_f$ a supplementary length that is different for the x and y axes: $2EL$ for the x axis and $2FL$ for the y axis. The $L_{tx} - L_{ty}$ difference may be quite significant. For example, for a TC with the parameters specified in Table 1, $L_{tx} - L_{ty} = 14.1$ m, i.e. for $k_{x,y} = 0.01k_0$ the difference in the time lag $\tau(k_x = 0.01k_0, k_y = 0) - \tau(k_x = 0, k_y = 0.01k_0)$ is 2.35ps. Note that the phase front curvature is proportional to $L_{sx,sy}$ and the intensity front curvature is proportional to $L_{tx,ty}$. Contrary to the case of free space propagation, these curvatures do not coincide after the TC. Moreover, the intensity front in y-direction may be concave because $F < 0$ as shown in Fig. 2. All other fronts are always convex because $C, D, E > 0$.

## V. AFGC AS A FILTER OF SPATIAL FREQUENCIES

The effects considered in Section 4 for TC are exactly the same as those in AFGC. The only difference is that the effective lengths (16, 18) have the form

$$L_{sx} = C_1L_1 + C_2L_2 + L_f \qquad\qquad L_{sy} = C_1L_1 + C_2L_2 + L_f \qquad (19)$$

$$L_{tx} = E_1L_1 + E_2L_2 + L_f \qquad\qquad L_{ty} = F_1L_1 + F_2L_2 + L_f, \qquad (20)$$

which directly follows from (10, 11, 12, 13). The most important distinctive feature of AFGC is that (8, 9) are not zeroed, and there appear in (7) two terms and, hence, two effects proportional to the first power of $k_x$: $\psi''_{x\omega}\Omega k_x$ that is the time lag proportional to $k_x$ or shift along the x axis proportional to $\Omega$ (spatial chirp), and also $\frac{1}{2}\psi'''_{x\omega\omega}\Omega^2 k_x$ that is GVD (pulse stretching) proportional to $k_x$ or shift along the x axis quadratic with respect to $\Omega$. As seen from (8, 9), the measure of compressor asymmetry are two parameters having the dimension of length:

$$\Delta L = A_1L_1 - A_2L_2 \qquad\qquad \Delta L' = B_1L_1 - B_2L_2$$

### V.1. Time lag proportional to $k_x$ or shift along the x axis proportional to $\Omega$ : $\psi''_{x\omega}\Omega k_x$

The physical meaning of this term becomes transparent if we write $\psi''_{x\omega}\Omega k_x = X(\Omega)\, k_x = \tau_A(k_x) \cdot \Omega$, where $X(\Omega) = \psi''_{x\omega}\Omega$ is the beam shift along the x axis (spatial chirp) and $\tau_A(k_x) = \psi''_{x\omega}k_x$ is the time lag. These two representations are equivalent and are illustrated in Fig. 2 for clarity. From the point of view of spatial noise, this effect reduces to temporal filtering, and it is convenient to consider it in terms of the pulse time lag with $k_x \neq 0$:

$$\tau_A(k_x) = -\frac{k_x}{k_0} \cdot \frac{\Delta L}{c} \qquad (21)$$

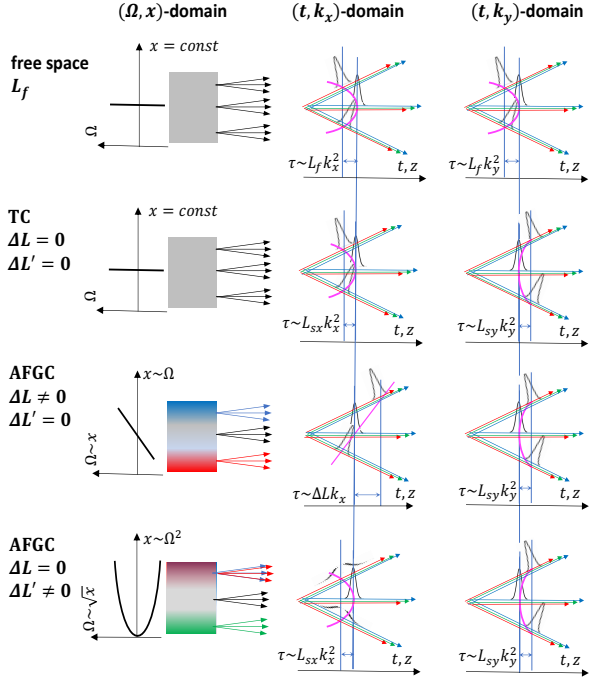

Fig. 2. Schematic representation of the field after propagation in free space, TC and AFGC in three domains. Solid magenta lines depict intensity front.

Unlike temporal self-filtering, the time lag here is proportional to the first rather that the second (17) power of $k_x$, which makes the effect much stronger. However, for $\Delta L \leqslant L_x \frac{k_x}{k_0}$, the effects are comparable. The second difference is that the sign of the lag depends on the sign of $k_x$, i.e. the spatial noise for which $k_x \Delta L > 0$, will *overtake* the main pulse rather lag behind it (Fig. 2). Lagging and overtaking make an identical impact on fluence smoothing. At the same time, the pulses propagating ahead of the main pulse reduce the time contrast. Note that this parasitic effect is especially strong in an SSGC. For instance, for a compressor with parameters listed in Table 1, even large-scale (usually highly energetic) noise with $k_x = 10^{-4} k_0$ overtake the main pulse by 0.77 ps. According to the contrary assertion made in [25], the integrated contrast does not degrade, which is evidently due to the neglect of the space-time coupling effects considered above.

### V.2. Time lag proportional to $k_x$ or shift along the x axis proportional to $\Omega$ : $\psi'''_{x\omega\omega}\Omega k_x$

The physical meaning of this term is quite obvious – pulses with a wave vector $k_x$ acquire, in addition to the principal wave with $k_x = 0$, a chirp (GVD) proportional to $k_x$:

$$\frac{1}{2}\psi'''_{x\omega\omega} k_x \Omega^2 = GVD(k_x) \cdot \Omega^2 = k_x \Delta L' \cdot \frac{\Omega^2}{\omega_0^2}$$

If the main pulse (with $k_x = 0$) at the compressor output is Fourier-transform-limited, then the noise with $k_x \neq 0$ will be extended in time independent of the sign of $k_x$ (Fig. 2). This reduces filtering caused by the noise time lag or overtaking, as stretching of the lagging or overtaking noise pulse increases its time overlap with the main pulse.

A proper choice of $L_{1,2}$, $N_{1,2}$ and $\alpha_{1,2}$ allows one to control values of $\Delta L$ and $\Delta L'$, thus creating complex space-time distributions of the field in the focal plane that may be useful for charged particle acceleration or other applications. In particular, the combinations $\Delta L' = 0$, $\Delta L \neq 0$ or $\Delta L' \neq 0$, $\Delta L = 0$ demonstrated in Fig. 2 may be implemented.

## VI. REDUCING FLUENCE FLUCTUATIONS AT THE AFGC OUTPUT

As mentioned above, all the effects proportional to the second power of $k_{x,y}$ in a TC as well as in an AFGC may be reduced to analogous effects in free space. The suppression of fluence fluctuations at free space propagation was analyzed analytically in [22, 23]. These earlier results may be generalized taking into account (10, 11, 12, 13). Here, we will not dwell on this, instead, to clearly demonstrate the effect of compressor asymmetry, we will focus on the effects proportional to the first power of $k_x$, specific for AFGC. To do this, in (7) we will neglect the last four terms proportional to $k_{x,y}^2$ and leave only the first two terms. Let us compare fluence fluctuations at the AFGC output with a reference, namely a TC that is a particular case of AFGC with $\Delta L = \Delta L' = 0$. In this case, the field $E_{ref}(t, x, y)$ at the TC output may be written as a sum of the main $E_0(x,y)U(t)$ and noise $E_{ref,n}(t,x,y)$ fields:

$$E_{ref}(t,x,y) = E_0(x,y)U(t) + E_0(x,y)U(t)f(x,y), \qquad (22)$$

where $f(x,y)$ is a complex function and $|f(x,y)| \ll 1$; $\int f(x,y) dS = 0$. From (3, 7, 8, 9) we obtain

$$E_{out}(\omega, k_x, k_y) = E_{ref}(\omega, k_x, k_y) \exp\left(i\left\{-\frac{\Delta L}{\omega_0}\Omega + \frac{\Delta L'}{\omega_0^2}\Omega^2\right\} k_x\right) \qquad (23)$$

The fluence $w_{out}(x,y)$ may also be represented as a sum of the main $w_{out,0}(x,y)$ and noise $w_{out,n}(x,y)$ fluences:

$$w_{out}(x,y) = \int |E_{out}(t,x,y)|^2 dt = w_{out,0}(x,y) + w_{out,n}(x,y) \qquad (24)$$

By substituting (22, 2) into (23) and the result into (24) we find $w_{out,n}(x,y)$, from which we obtain an expression for the spectrum of fluence fluctuations $w_{n,out}(k_x, k_y)$:

$$\left|w_{out,n}(k_x, k_y)\right|^2 = T(k_x) \cdot \left|w_{ref,n}(k_x, k_y)\right|^2, \qquad (25)$$

where

$$T(k_x) = \left|\frac{\int |U(\Omega)|^2 \cdot \exp\left(i(-\frac{\Delta L}{\omega_0}\Omega + \frac{\Delta L'}{\omega_0^2}\Omega^2)k_x\right) d\Omega}{\int |U(\Omega)|^2 d\Omega}\right|^2 \qquad (26)$$

$T(k_x)$ has the meaning of the spectrum filter transmittance. Two terms in the exponent correspond to two effects – lag (overtaking) and GVD. If we neglect the latter, i.e. if $\frac{\Omega}{\omega_0} \ll \frac{\Delta L}{\Delta L'}$, then from (26) we obtain a result fully analogous to temporal self-filtering [22, 23]: the filter transmittance is equal to the square of the modulus of the autocorrelation function of the field $A_t$ with the argument equal to the time lag $\tau_A$ (21):

$$\left|w_{out,n}(k_x, k_y)\right|^2 = |A_t(\tau_A)|^2 \cdot \left|w_{ref,n}(k_x, k_y)\right|^2, \qquad (27)$$

where

$$A_t(\tau) = \frac{\int U(t)U^*(t-\tau) dt}{\int |U(t)|^2 dt} \qquad (28)$$

The difference from the temporal self-filtering is that the time lag $\tau_A$ (21), as distinct from $\tau$ (17), is proportional to the first (rather than the second) power of $k_x$, besides, $\tau_A$ does not depend on $k_y$. As a reference noise spectrum $w_{ref,n}(k_x, k_y)$ we take the frequently used model [28, 29]

$$\left|w_{ref,n}(k_x, k_y)\right|^2 = \frac{const}{(h^2 + k_x^2 + k_y^2)^\gamma}, \qquad (29)$$

where $\gamma$ is the constant that takes on a value from 1 to 2 (usually 1.28), and the spatial scale $\Lambda = 2\pi/h$ is smaller but commensurable with the beam size. By integrating (27) with respect to $k_y$, with (29) taken into account, we obtain for a Gaussian pulse $U(t) = \exp(-(t^2/\tau_p^2))$ a one-dimensional spectrum

$$S(k_x) = \int_{-\infty}^{\infty} |w_{out,n}(k_x, k_y)|^2 k_y = const \cdot \frac{B\left(\frac{1}{2}; \gamma - \frac{1}{2}\right)}{(h^2 + k_x^2)^{\gamma - \frac{1}{2}}} \cdot \exp\left(-\left(\frac{\Delta L}{c\tau_p} \cdot \frac{k_x}{k_0}\right)^2\right), \quad (30)$$

where $B(a; b)$ is a B-function. The $S(k_x)$ spectrum for $\tau_p = 30$ fs is plotted in Fig. 3 at different values of $\Delta L$. Figure 3a clearly demonstrates filtering of high spatial frequencies in AFGC (cf. the colored and the black curves). Knowing the fluence fluctuation spectrum $|w_n(k_x, k_y)|^2$, one can find the rms of the fluctuations:

$$rms = \sigma_{out,ref} = \sqrt{\int_{-\infty}^{\infty} |w_{n;ref,out}(k_x, k_y)|^2 dk_x dk_y}, \quad (31)$$

that is a convenient quantitative characteristic of filtering efficiency: the $\sigma_{out}/\sigma_{ref}$ ratio shows how many times the fluence fluctuation rms will decrease by AFGC compared to TC. The substitution of (30) into (31) and integration with respect to $dk_x$ yields

$$\left(\frac{\sigma_{out}}{\sigma_{ref}}\right)^2 = \frac{2}{B\left(\frac{1}{2}; \gamma - 1\right)} \int_0^\infty \frac{\exp\left(-\left(\frac{\lambda}{\Lambda} \frac{\Delta L}{c\tau_p} \cdot x\right)^2\right)}{(1+x^2)^{\gamma - 1/2}} dx \quad (32)$$

For a given $\gamma$, the $\frac{\sigma_{out}}{\sigma_{ref}}$ ratio depends only on one parameter $\frac{\lambda}{\Lambda} \frac{\Delta L}{c\tau_p}$. The corresponding dependences are plotted in Fig. 4. One can see from the figure that the value of $\gamma$ is not very important and at large $\frac{\lambda}{\Lambda} \frac{\Delta L}{c\tau_p}$ the value of $\frac{\sigma_{out}}{\sigma_{ref}}$ is proportional to $\sqrt{\frac{\lambda}{\Lambda} \frac{\Delta L}{c\tau_p}}$. Note that for beams of a smaller diameter (smaller $\Lambda$), the high-frequency components have larger values and, hence, filtering is more efficient. It is clear from (32) and Figs. 3,4 that filtering efficiency is determined by the ratio of time $\Delta L/c$ to the pulse duration at the compressor output $\tau_p$, which once again emphasizes the relevance of the used term "temporal filtering".

To conclude we note that to find the probability for the fluence to exceed the threshold value, it is sufficient to know the fluctuation spectrum $|w_n(k_x, k_y)|^2$ [18].

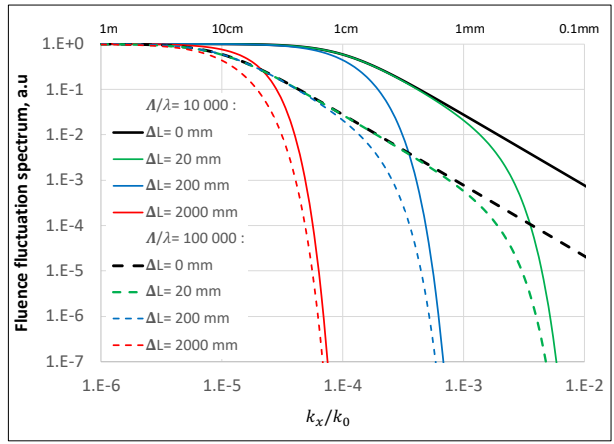

Fig. 3 One-dimensional fluence fluctuation spectrum $S(k_x)$ at AFGC output for a pulse having duration $\tau_p = 30$ fs ($\gamma = 1.28$). The difference between the colored and black curves shows the efficiency of fluence smoothing at a given $k_x$. The efficiency of SSSF suppression for a pulse with duration $\tau_p = 21.2$ fs will be equal to the efficiency of fluence smoothing $S(k_x)$ (see the text).

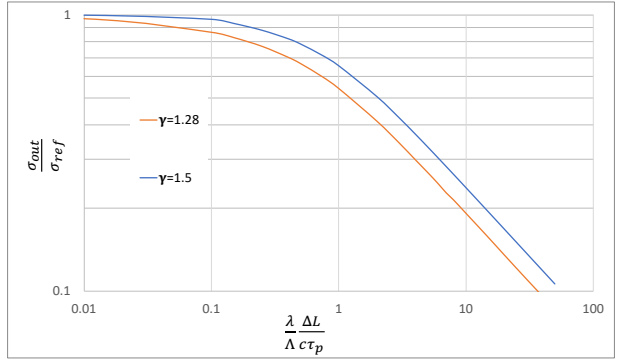

Fig. 4. Reducing the rms of fluence fluctuation $\frac{\sigma_{out}}{\sigma_{ref}}$.

## VII. SUPPRESSING SMALL-SCALE SELF-FOCUSING BY MEANS OF AFGC

Exact computation of the efficiency of small-scale self-focusing demands solution of a nonlinear nonstationary Schrödinger equation in which the field at the compressor output (4) is used as a boundary condition at z=0. This requires numerical simulation, which is outside the scope of the present work. However, the results obtained allow us to assess the efficiency of temporal filtering. For suppressing small-scale self-focusing it is necessary to reduce noise *intensity* $I_n$ (rather than noise fluence) at the time of maximum main pulse, i.e. at the time of maximum nonlinearity. Consequently, the reduction of $I_{out,n}(t = 0, k_x, k_y)$ will be a good estimate. After the corresponding computations for $I_{out,n}(t = 0, k_x, k_y)$, similarly to Section 6 we will obtain an expression analogous to (27):

$$I_{out,n}(t = 0, k_x, k_y) = \left|U\left(t = -\frac{k_x}{k_0}\frac{\Delta L}{c}\right)\right|^2 I_{ref,n}(t = 0, k_x, k_y) \qquad (33)$$

Thus, from the point of view of suppressing small-scale self-focusing, an AFGC is a filter with the transmittance equal to the normalized pulse intensity at the time $-\frac{k_x}{k_0}\frac{\Delta L}{c}$. For a Gaussian pulse, $|U(t)|^2 = \left|A_t(t/\sqrt{2})\right|^2$. Therefore, with (27, 33) taken into account, filtering efficiency for the intensity will be the same as for the fluence shown in Fig. 3, but for a $\sqrt{2}$ times shorter pulse duration, i.e. 21.2 fs instead of 30 fs. As mentioned above, of primary importance is the reduction of the value of the noise intensity spectrum $I_n(t = 0, k_x, k_y)$ at spatial frequencies $k_{max}$, for which the instability increment is the largest. For high-power femtosecond lasers, $k_{max}/k_0$ is of the order of 0.03 [4, 5], i.e. as seen in Fig. 3, a small compressor asymmetry $\Delta L$ is quite sufficient for effective filtering.

## VIII. CONCLUSION

For reducing beam fluence fluctuations it was proposed [1] to use a compressor in which diffraction gratings in two pairs are different, whereas all the other parameters are the same. However, this effect was investigated only numerically. In the presented paper, an analytical theory has been constructed that describes all space-time coupling effects arising in an asymmetric optical compressor, in which pairs of gratings may differ not only by the distance between the gratings but also by the groove density and angle of incidence. It has been shown that no compressor asymmetry affects the far field fluence and on-axis focal intensity. Given that the gratings in each pair are parallel and identical, this conclusion is true for any compressor, including the one with a single grating pair.

From the point of view of beam cleanup from spatial noise, a symmetric compressor is "similar" to free space in that it also accomplishes spatial and temporal self-filtering of a beam. The difference is that both these effects become anisotropic: the compressor is an anisotropic transmissive filter. In particular, it introduces astigmatism.

In an asymmetric compressor, there appear two additional effects proportional to the first power of the transverse wave vector, that are shown schematically in Fig. 2. The spatial noise i) lags behind/overtakes the main pulse, which is equivalent to a linear spatial chirp, and ii) acquires a temporal chirp (GVD), which is equivalent to a squared spatial chirp. The first effect reduces beam fluence fluctuations at the compressor output. Exact expressions for the fluence fluctuation spectrum and fluence rms have been obtained, with the latter being dependent only on one parameter characterizing compressor asymmetry. The second effect reduces filtering induced by the first effect, as noise pulse stretching increases its time overlap with the main pulse. By choosing adequate grating parameters it is possible to control these two effects independently, e.g., for creating complex space-time field distributions in the focal plane that may be interesting for different applications.

The asymmetric compressor is also interesting for suppressing small-scale self-focusing, e.g., at subsequent post-compression. In this case, it is necessary to reduce the noise intensity rather than its fluence, which should be done at high spatial frequencies. The constructed theory made it possible to estimate the efficiency of suppression of small-scale self-focusing and showed that it may be orders of magnitude, even with a slight asymmetry of the compressor.

The disadvantages of an asymmetric compressor include the degradation of temporal contrast. The theory of this inevitable parasitic effect will be considered elsewhere.

## APPENDIX. PHASE INCURSION IN A COMPRESSOR FROM POINT D1 TO POINT D2

As a pulse is propagating from point A1 to point C1 (Fig. 1) there occurs phase incursion

$$\Psi_1 = L_1 k_{zx} \left( cos\theta_1 + cos\left\{ \alpha_1 + atan\frac{k_x}{k_z} \right\} \right), \quad (34)$$

where $L_1$ is the distance between the gratings along the normal, $k_{zx}^2 = \left(\frac{\omega}{c}\right)^2 - k_y^2$, and $\theta$ is found from the grating equation

$$sin\theta_1(\omega, k_x) = m \frac{2\pi}{k_{zx}(\omega)} N_1 + sin\left\{ \alpha_1 + atan\frac{k_x}{k_z(\omega)} \right\} \quad (35)$$

The choice of the position of points A1 and C1 on the z axis is arbitrary, but it is important that they should have the same coordinates on the z axis, i.e. C1 must be strictly above A1. The expression (34) was derived in the classical work by Treacy [27] in other notation. Note that sometimes (see, e.g., [25]) instead of (34) it is assumed that $\Psi_1 = \frac{\omega}{c} p$, where $p$ is the path length from A1 to C1, which contradicts (34). Taking into account that $k_{zx} \approx \frac{\omega}{c}\left(1 - \frac{k_y^2}{2(\omega/c)^2}\right)$, from (34, 35) to an accuracy of $k_{x,y}^3$ we obtain

$$\Psi_1(\omega, k_x, k_y) = L_1 \frac{\omega}{c}\left( cos\theta_1 + cos\left\{ \alpha_1 + \frac{k_x c}{\omega} \right\} \right) - L_1 \frac{c}{2\omega} k_y^2 \left( cos\{\theta_1(k_{x,y} = 0)\} + cos\alpha_1 \right) \quad (36)$$

$$sin\theta_1(\omega, k_x, k_y) = m \frac{2\pi c}{\omega} N_1 \left( 1 + \frac{c^2}{2\omega^2} k_y^2 \right) + sin\left\{ \alpha_1 + \frac{k_x c}{\omega} \right\} \quad (37)$$

The second pair of gratings from point A2 to point C2 is equivalent to the first pair from point A1 to C1 accurate to the replacement of $\Delta\alpha$ by $-\Delta\alpha$ or $k_x$ by $-k_x$, since for $\Delta\alpha > 0$, $\alpha_1$ increases but $\alpha_2$ decreases, which is well seen in Fig. 1. In other words, the x axis changes the direction (the x axis is always directed outward from the grating, rather than inward), and the vector $\mathbf{k}_x$ does not change direction, as the gratings are parallel. Thus, from point A2 to point C2, there is spectral phase incursion $\Psi_2 = \Psi_1(-k_x)$, with all subscripts "1" in the expression for $\Psi_1$ (36) replaced by "2".

During pulse propagation from the input point D1 to the output point D2 the phase incursion is $\Psi = \Psi_1 + \Psi_2 + \Psi_f$, where $\Psi_f$ is the phase incursion in free space of length $L_f$:

$$\Psi_f(\omega, k_x, k_y) = L_f \frac{\omega}{c} - L_f \frac{c}{2\omega}(k_x^2 + k_y^2), \tag{38}$$

where $L_f = |D1A1| + |C1A2| + |C2D2|$. As seen from Fig. 1, $L_f$ is the distance between the input and output planes of the compressor. The choice of these planes is arbitrary but it is important that the distance between them is $L_f$. Finally, from (36, 37, 38) we obtain the expressions (5, 6):

$$\Psi(\omega, k_x, k_y) = L_1 \frac{\omega}{c}\left(\cos\theta_1 + \cos\left\{\alpha_1 + \frac{k_x c}{\omega}\right\}\right) + L_2 \frac{\omega}{c}\left(\cos\theta_2 + \cos\left\{\alpha_2 - \frac{k_x c}{\omega}\right\}\right) + L_f \frac{\omega}{c} - \frac{c}{2\omega} k_y^2 [L_f + (\cos\{\theta_1(k_{x,y}=0)\} + \cos\alpha_1)L_1 + (\cos\{\theta_2(k_{x,y}=0)\} + \cos\alpha_2)L_2]$$

$$\sin\theta_{1,2}(\omega, k_x, k_y) = m\frac{2\pi c}{\omega} N_{1,2}\left(1 + \frac{c^2}{2\omega^2} k_y^2\right) + \sin\left\{\alpha_{1,2} \pm \frac{k_x c}{\omega}\right\}$$